\documentclass[3p,times]{elsarticle}

\usepackage{ecrc}
\usepackage[final]{epsfig}

\volume{00}

\firstpage{1}

\journalname{Nuclear Physics A}

\runauth{}

\jid{npa}

\jnltitlelogo{Nuclear Physics A}

\usepackage{amssymb}

\biboptions{square,comma,numbers,sort&compress}

\usepackage[figuresright]{rotating}

\begin{document}

\begin{frontmatter}



\dochead{}

\title{Jet correlations --- opportunities and pitfalls}


\author{Thorsten Renk}

\address{Department of Physics, P.O. Box 35, FI-40014 University of Jyv\"askyl\"a, Finland and \\Helsinki Institute of Physics, P.O. Box 64, FI-00014 University of Helsinki, Finland}

\begin{abstract}
The simplest observables used to probe the interaction of hard partons with a QCD medium in ultrarelativistic heavy ion  collisions measure disappearance, such as the nuclear modification factor $R_{AA}$. The information content of such observables is however limited. More differential information is obtained from triggered correlation observables where a trigger condition ensures that a hard event has taken place and the correlation of other objects in the event with the trigger contains information about the nature of parton-medium interaction. By construction, triggered correlation observables are conditional probabilities, i.e. they measure events biased by the trigger condition. The presence of this bias makes the interpretation of observables non-intuitive, but  at the same time represents an opportunity to design future measurements to selectively probe particular physics. In this work, an overview over the four types of biases occuring in triggered hard correlation observables is given, followed by a study of current jet correlation phenomenology in the light of the preceding discussion.
\end{abstract}


\end{frontmatter}


\section{Jet correlations in theory and experiment}

The idea underlying the use of hard probes in the context of ultrarelativistic heavy-ion (A-A) collisions is to gain information on both macroscopic (i.e. geometry and evolution) and microscopic (i.e. relevant degrees of freedom) of the droplet of QCD matter produced in such collisions. This is expected to work since the uncertainty relation can be used to argue that the hard process itself will take place without influence by the surrounding medium, leaving the interaction with the medium as a final state effect modifying a perturbatively calculable process. In other words, the attenuation pattern of hard partons propagating through the medium can be used to do tomography since the initial production rate of these partons is under control.

\begin{figure}[htb]
\begin{center}
\epsfig{file=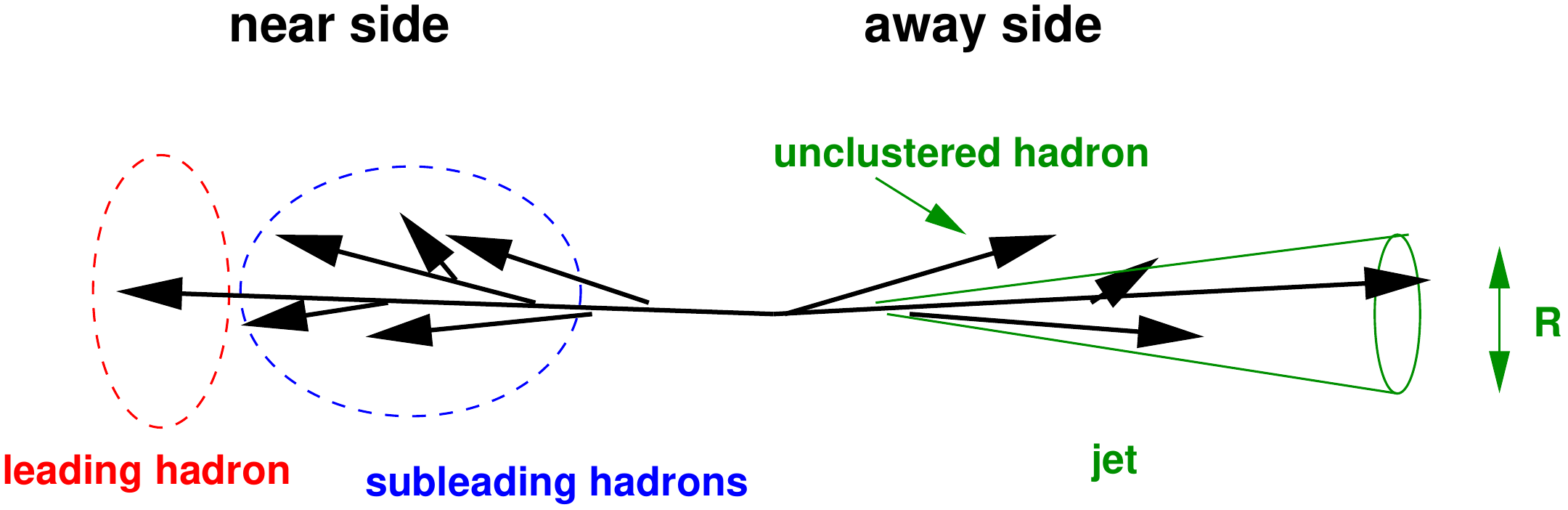, width=9cm}
\end{center}
\caption{\label{F-1}Sketch of various ways of observing a hard back-to-back event experimentally}
\end{figure}

The simplest class of measurements focuses on disappearance, such as the the nuclear modification factor $R_{AA}$ for single hadrons or reconstructed jets. These reveal that the interaction with the medium suppresses high $P_T$ hadrons and jets \cite{PHENIX-RAA,ALICE-RAA,CMS-RAA}, but not much about the underlying mechanism. In order to probe the physics more differentially, other observables offer themselves, cf. Fig.~\ref{F-1}.

The basic topology of a hard event consists of two highly virtual partons, approximately back to back, which evolve first in terms of a parton shower while passing through the medium before they hadronize. However, this process needs to be detected and studied on the background of hadrons coming from the medium itself. For this purpose, typically a trigger condition is evaluated to make sure that the transvserse momentum $P_T$ scale of the process is above some threshold. The trigger condition can refer to a single hadron (typically the leading hadron of the shower) or a clustering algorithm \cite{FastJet} can be used to combine part of the collimated spray of hadrons into jets. Given a valid trigger, other hadrons in the event can now be correlated and analyzed, both on the trigger (near) side and on the away side, which defines the range of experimentally available correlations (h-h, jet-h, h-jet, \dots). Note that also analyses of properties of jets such as the CMS jet shape analysis \cite{CMS-Jetshape} are technically a triggered correlation analysis and hence a conditional probability --- first a jet in the given energy range is found, then given the trigger the near side hadron distribution is evaluated.  It is thus important to understand the bias imposed by the respective trigger condition before interpreting any results.

\begin{figure}[htb]
\begin{center}
\epsfig{file=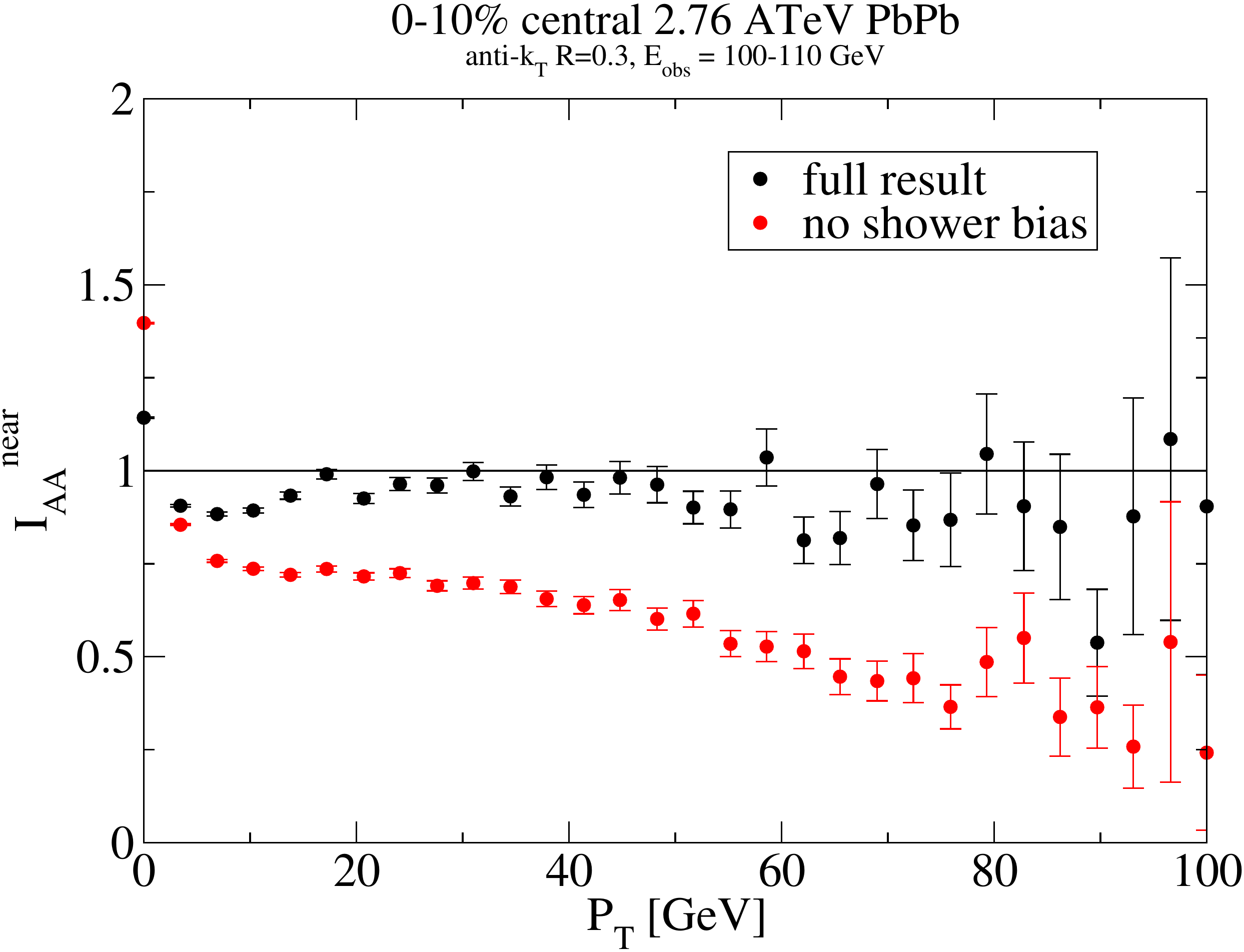, width=7cm}\epsfig{file=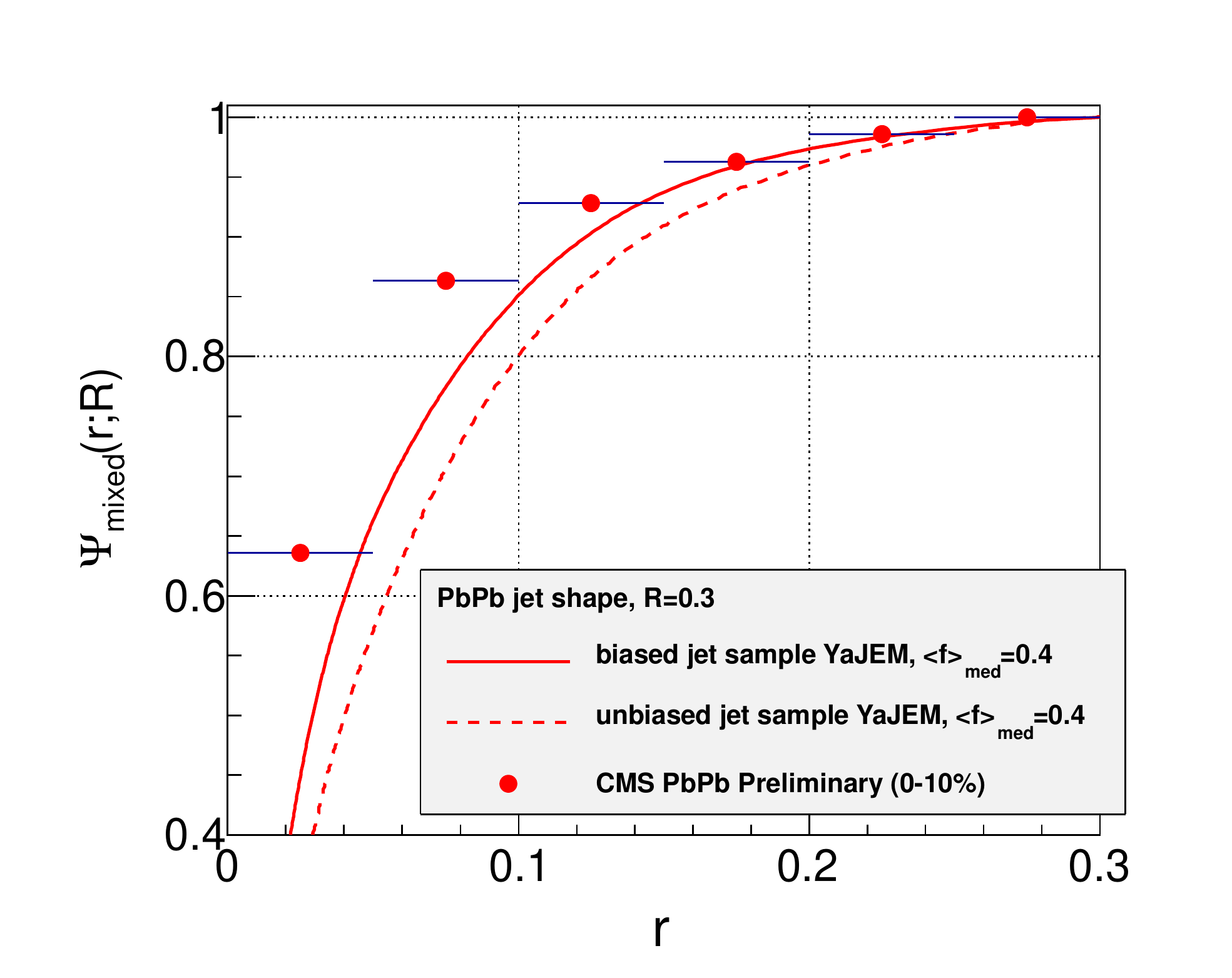, width=7cm}
\end{center}
\caption{\label{F-2}Left panel: Ratio of medium over vacuum fragmentation function for 100 -110 GeV jets, comparing the result without jet finding bias with the full result \cite{Bayesian} Right panel: Jet shape for 100 GeV jets, comparing biased vs. unbiased result \cite{Shape}. }
\end{figure}

This is in particular crucial in view of the fact that theoretical computations are often done forward in time, i.e. they start with a parton with a defined energy, propagate it through a specified medium and obtain a medium-modified hadron shower which can be clustered into a jet. In contrast, the experimental procedure clusters a final state to a defined energy and one has to conclude backwards in time what the original parton properties might have been. The existence of biases implies that there is no meaningful comparison between the two procedures, the only way to compare theory with experiment is to compute for all possible initial states and simulate how the experimental procedure selects out a particular subgroup of the resulting final states. As Fig.~\ref{F-2} shows, this is not an academic issue --- results with and without bias are qualitatively and quantitatively different. In short, a 100 GeV parton (or any fixed energy parton) is not a meaningful representative of an experimentally identified 100 GeV jet. 

\section{Types of biases}

Following the classification in \cite{Bayesian}, one can distinguish four main types of biases:

\begin{itemize}
\item The {\bfseries kinematic bias} is a medium-induced shift in the relation between trigger and underlying parton kinematics. Even in vacuum, the relation between the momentum of a triggered hadron or jet to the underlying parton is probabilistic, but due to the medium-induced modification to the shower evolution, this relation changes in medium such that the trigger contains on average a lower fraction of the parton momentum than in vacuum, i.e. for fixed trigger momentum, the underlying parton momentum \emph{increases} on average in medium.
\item The {\bfseries parton type bias} is a shift in the mixture of observed quark to gluon jets fulfilling a trigger condition. In vacuum, quarks have a harder fragmentation pattern and hence are more likely to fulfill a trigger condition. Since gluons couple by a factor $C_F = 9/4$ more strongly to the medium than quarks, the mixture changes in medium towards even fewer gluon-induced jets. if $qg \rightarrow qg$ happens to be an important channel, the bias enhances the gluon fraction on the away side however.
\item The {\bfseries geometry bias}  accounts for the fact that whenever the medium modification of a shower increases with medium density and traversed length, hard processes occuring close to the surface with a parton propagating outward are more likely to fulfill the trigger condition than partons which have to travserse the whole medium. The distribution of hard vertices leading to a trigger condition is hence almost never a binary overlap distribution but shifted towards a surface.  
\item Finally, while the previous three biases affect the structure of the back-to-back partonic event, the {\bfseries shower bias} is only relevant for the near side on which the trigger is obtained. Due to the trigger condition, the near side shower can no longer probe all possible configurations but is restricted, for instance triggering on a 20 GeV hadron, showers which contain only hadrons below 20 GeV can never be observed.
\end{itemize}

In practice, these biases occur intertwined and are not easily disentangled, with a strength that is largely determined by the precise choice of the trigger object, the underlying parton spectra as a function of $\sqrt{s}$ and the medium density.

As an example, consider a case study \cite {Bayesian} in which the away side conditional yield ratio ($I_{AA}$) is observed for a 12-15 GeV trigger object, where the trigger object may be either a photon, a hadron, an ideal jet resembling the CMS flow algorithm (clustering all hadrons in a radius of $R=0.4$ with the anti-$k_T$ algorithm) or a jet definition as has been used by the STAR collaboration  (clustering only particles above 2 GeV subject to PID cuts inside $R=0.4$).

\begin{figure}[htb]
\begin{center}
\epsfig{file=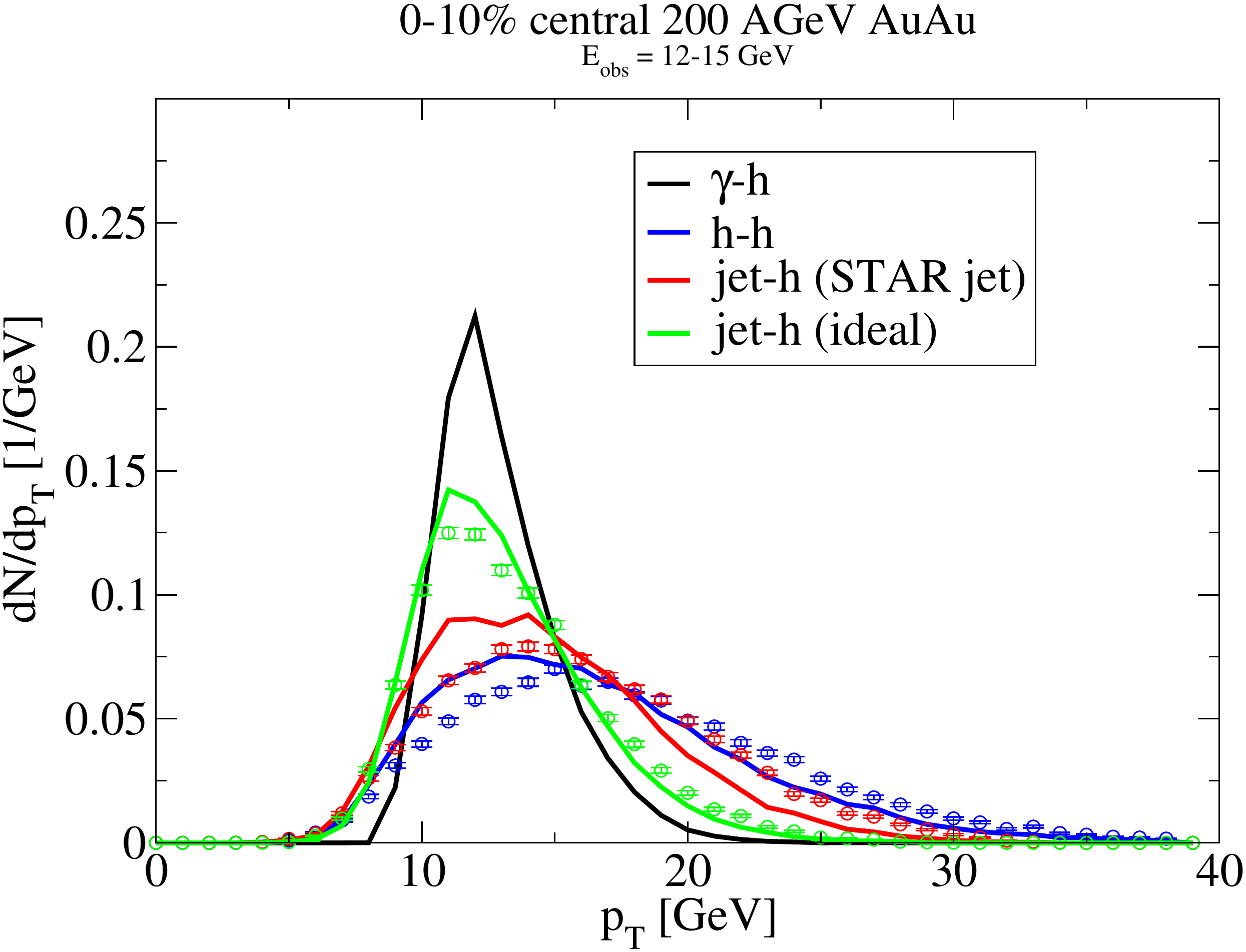, width=7cm}\epsfig{file=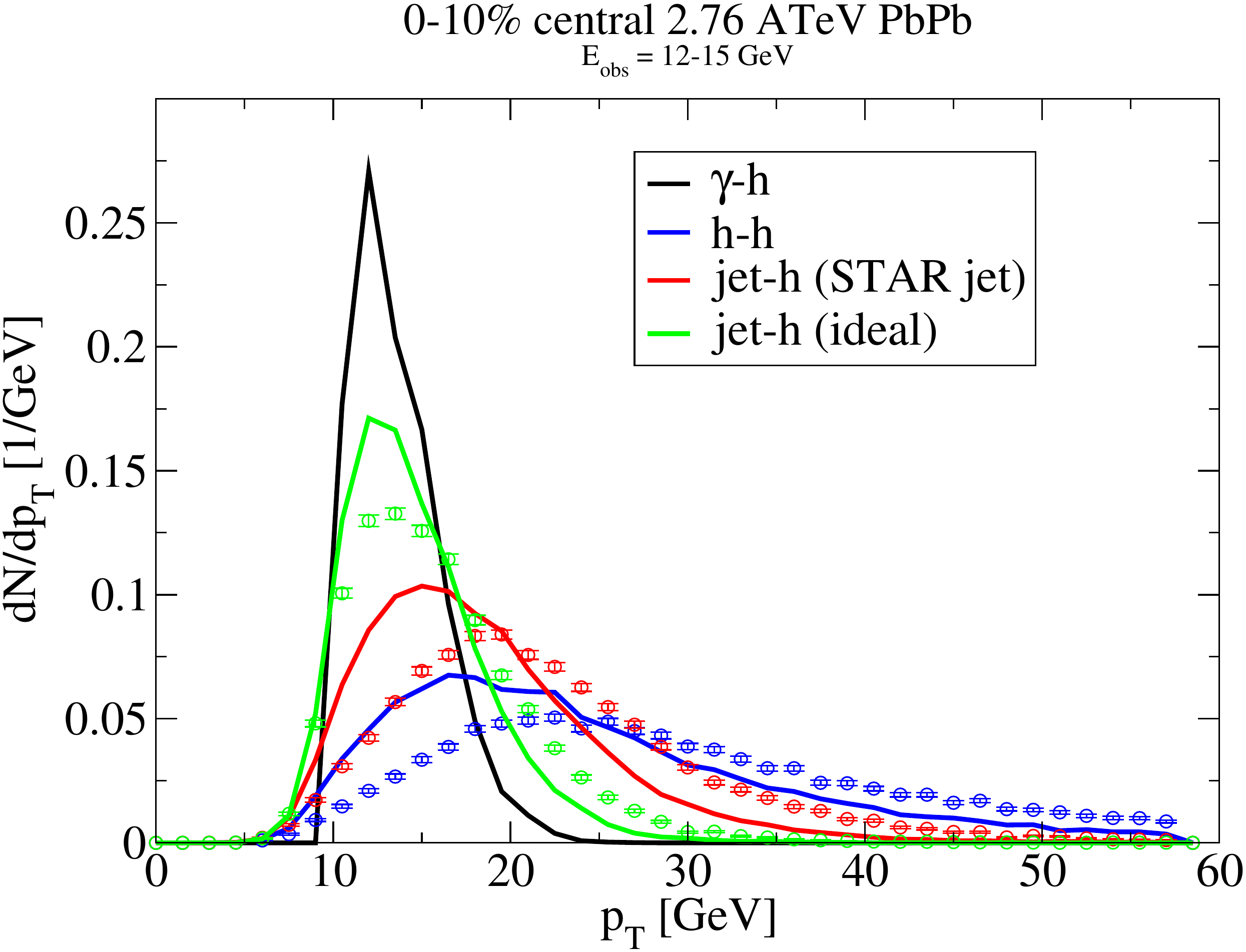, width=7cm}
\caption{\label{F-3} Away side parton momentum given a trigger object in the range of 12-15 GeV, lines are for vacuum, points show the situation in-medium. Left panel: RHIC kinematics, right panel: LHC kinematics}
\end{center}
\end{figure}

\begin{figure}[htb]
\begin{center}
\epsfig{file=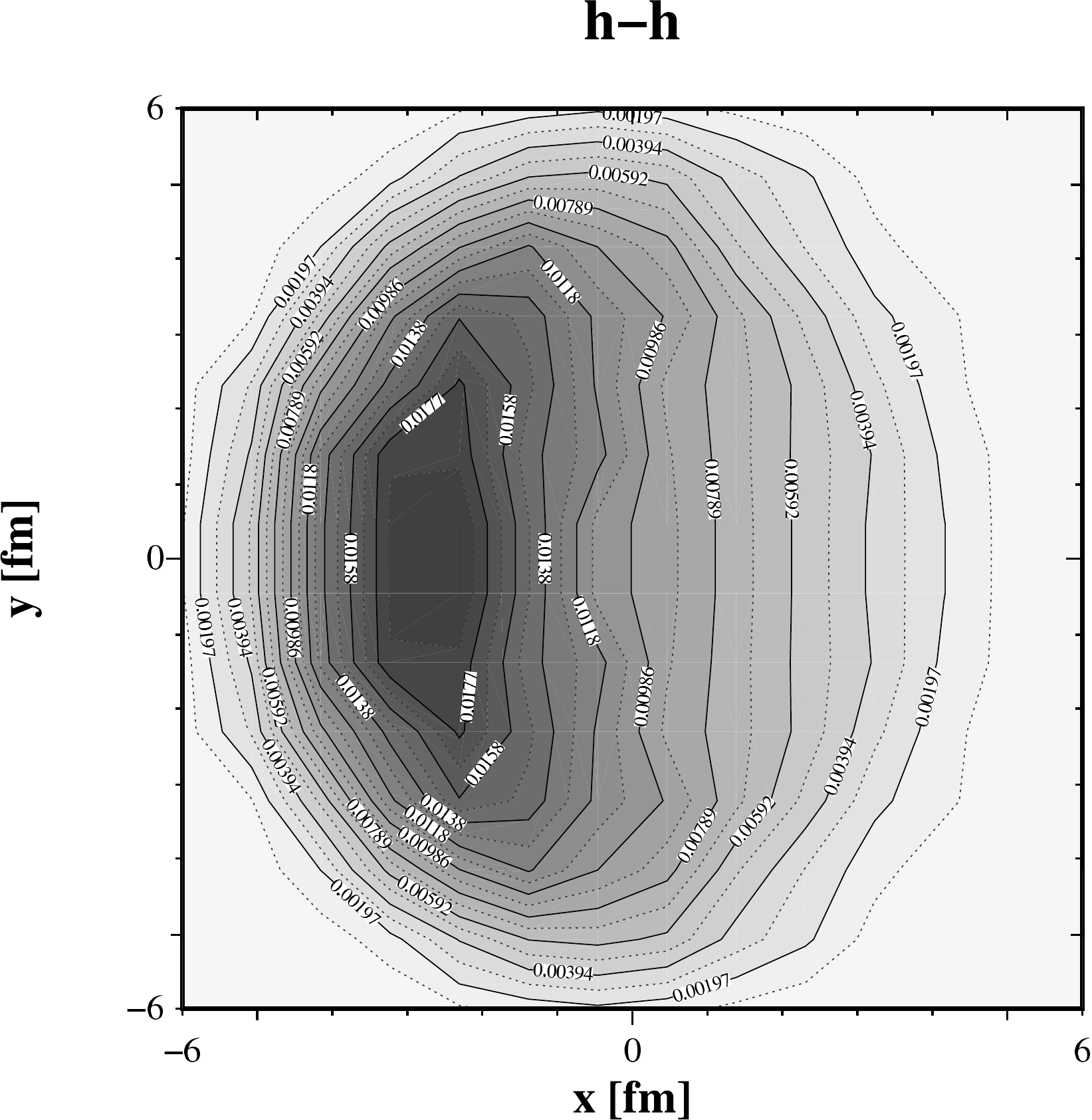, width=5cm}\epsfig{file=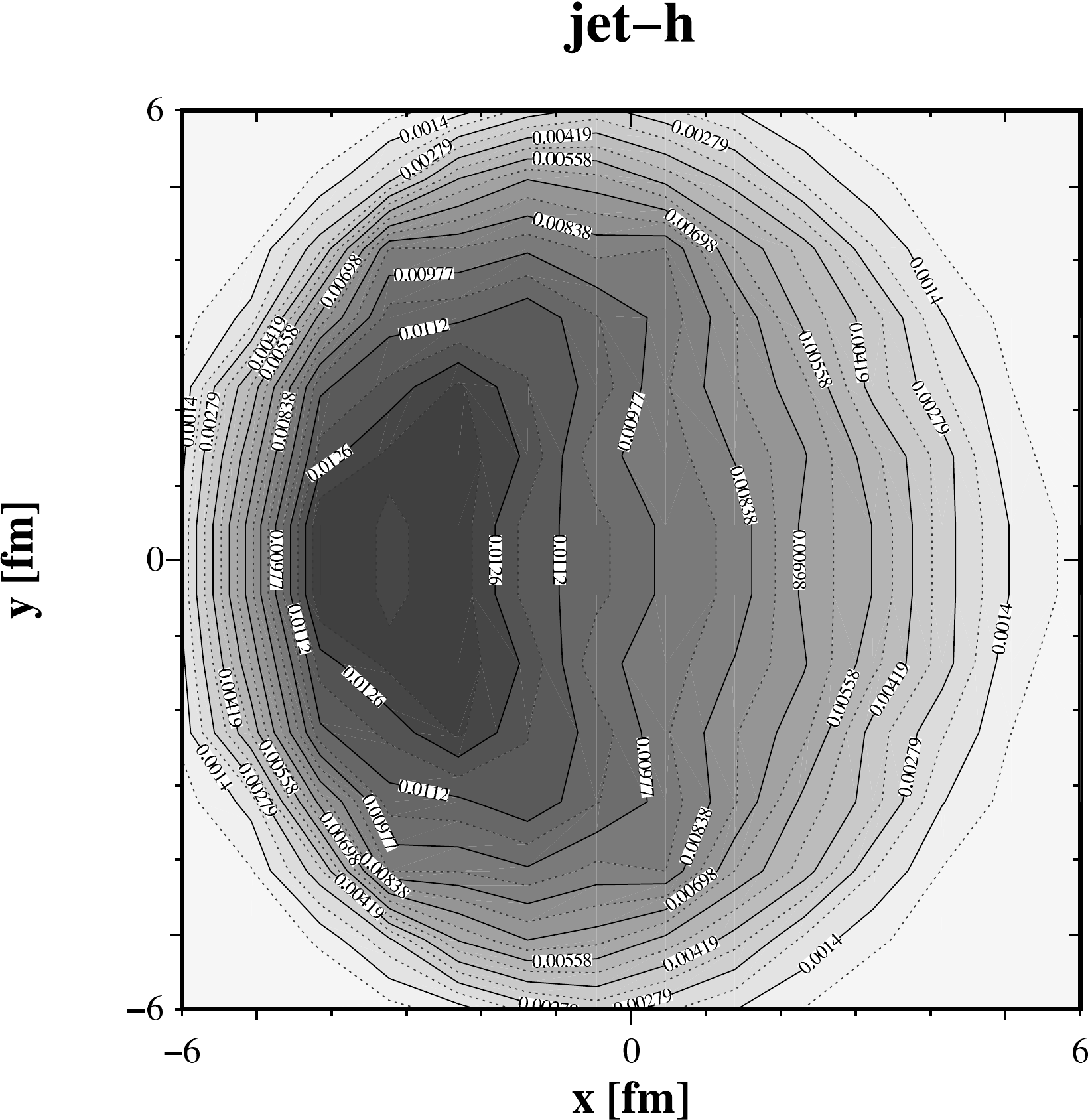, width=5cm}\epsfig{file=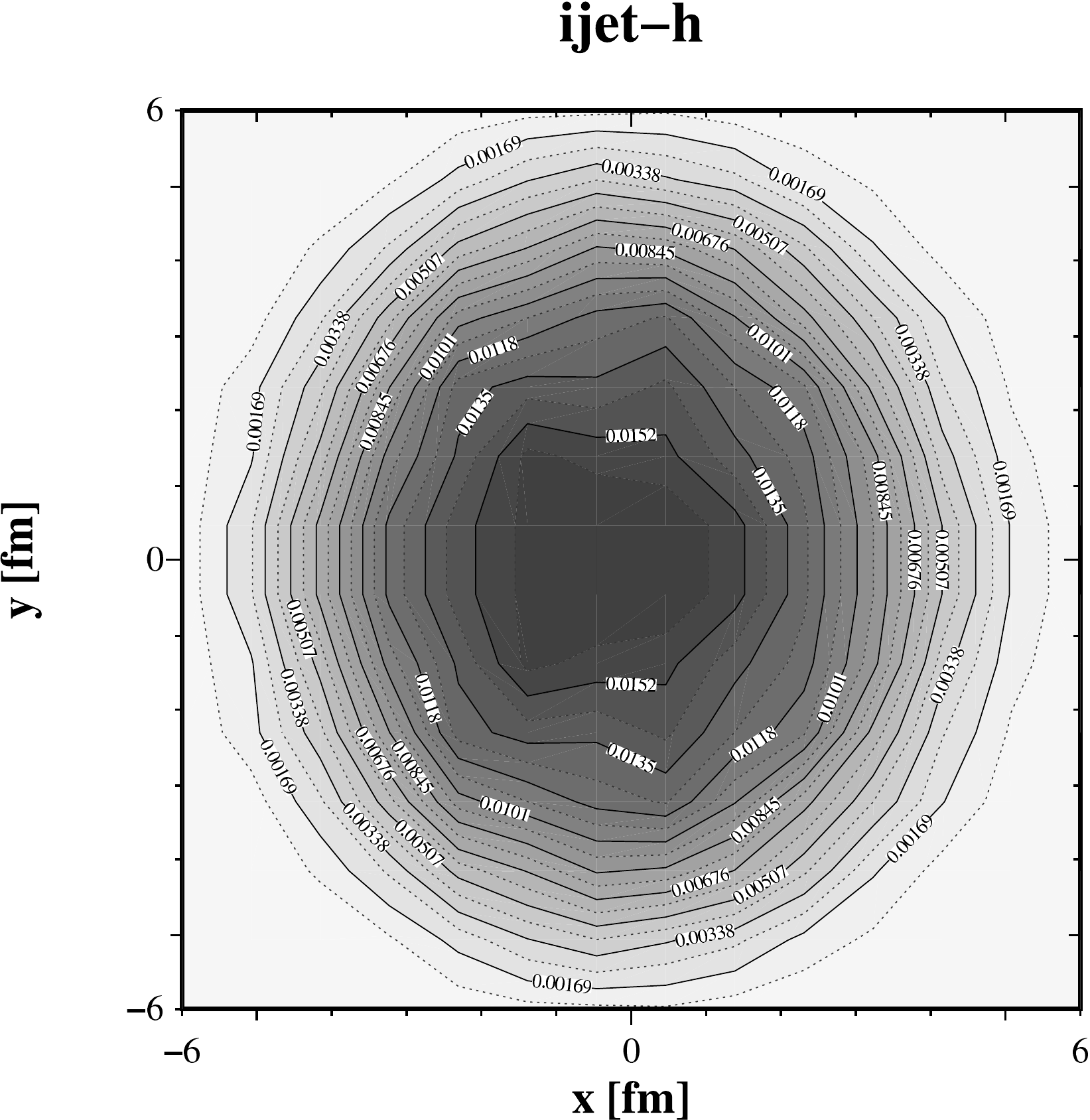, width=5cm}
\caption{\label{F-4} Distribution of vertices leading to a triggered event with the near side being the $-x$ direction for different trigger objects at RHIC kinematics. }
\end{center}
\end{figure}

Fig.~\ref{F-3} shows the kinematic bias, i.e. how the probabilistic relation between trigger and parton kinematics is shifted by the medium for different trigger objects and kinematics, whereas Fig.~\ref{F-4} shows the geometrical bias.

A number of conclusions can be drawn immediately. Fist, the details of the jet definition matter, jet definitions used by various experiments lead to visibly different biases and are hence not comparable. Second, if one aims at comparing jet corelations with different triggers, one should do so at comparable parton kinematics. Due to the different kinematic bias, comparing different trigger objects for the same trigger kinematics probes in fact vastly different partonic regimes. Third, also the hardness of the primary parton spectrum is important, kinematic shifts at LHC are visibly different and the geometry bias is significantly lessened (not shown here) due to the copious availability of hard partons as compared to RHIC. 

\section{A proposal}

As argued before, any observable hard probe represents an average over many theoretically possible initial state configurations. The results of the previous section suggest that the experimental trigger and analysis cut define the filter which controls this averaging process, and that this filter is in fact highly controllable: At RHIC kinematics, it is for instance conceptually possible to utilize highly surface-biased triggers like single hadrons and compare with minimally biased jet definitions. As Fig.~\ref{F-3} suggests, the average kinematics shift is largely understood in vacuum and the medium component can be approximately accounted for, thus one could for instance engineer to probe different regions of the medium at the same parton kinematics by suitable selections of different trigger objects and ranges, allowing position-differential medium tomography.

Thus, the aim should not be do make bias-free measurements (which, even if it were possible, might not be so interesting) but to understand and control the bias of a correlation measurement in such a way that it is particularly sensitive to a physics question. The same theoretical (and unobservable) object, i.e. a medium-modified shower, would in this way be studied through multiple filters, each filter being sensitive to a particular aspect of the problem. In this way, the information loss due to the need to average over a large number of possible initial states might be partially undone.

\section{Clustered vs. unclustered observables}

In Figs.~\ref{F-3} and \ref{F-4} we have seen that clustered (jet) triggers and unclustered (hadron) triggers lead to different biases. Whether one is more desirable than the other depends on what physics question one wants to study and other factors such as the rate of triggered events that can be achieved. The same is not true for the question whether the conditional observable should be a clustered or unclustered quantity --- here, generically unclustered observables retain more sensitivity to the physics of medium modification than clustered ones.

The reason is that the original purpose of clustering in $e^+e^-$ or p-p physics was to define a viable proxy for the hard, perturbative physics and to get rid of the complications of non-perturbative physics close to $\Lambda_{QCD}$ such as copious soft gluon emission and hadronization.  As a result, jet definitions are designed such that a jet is a reasonable representation of the initial parton kinematics. The implication is that clustering procedures are designed to suppress physics at a scale of $\Lambda_{QCD} \approx 300$ MeV --- which also happens to be the characteristic temperature scale of the medium. Since clustering is designed to undo the softer branching processes in a shower, it also tends to undo the medium modifications to this shower. 

\begin{figure}[htb]
\begin{center}
\epsfig{file=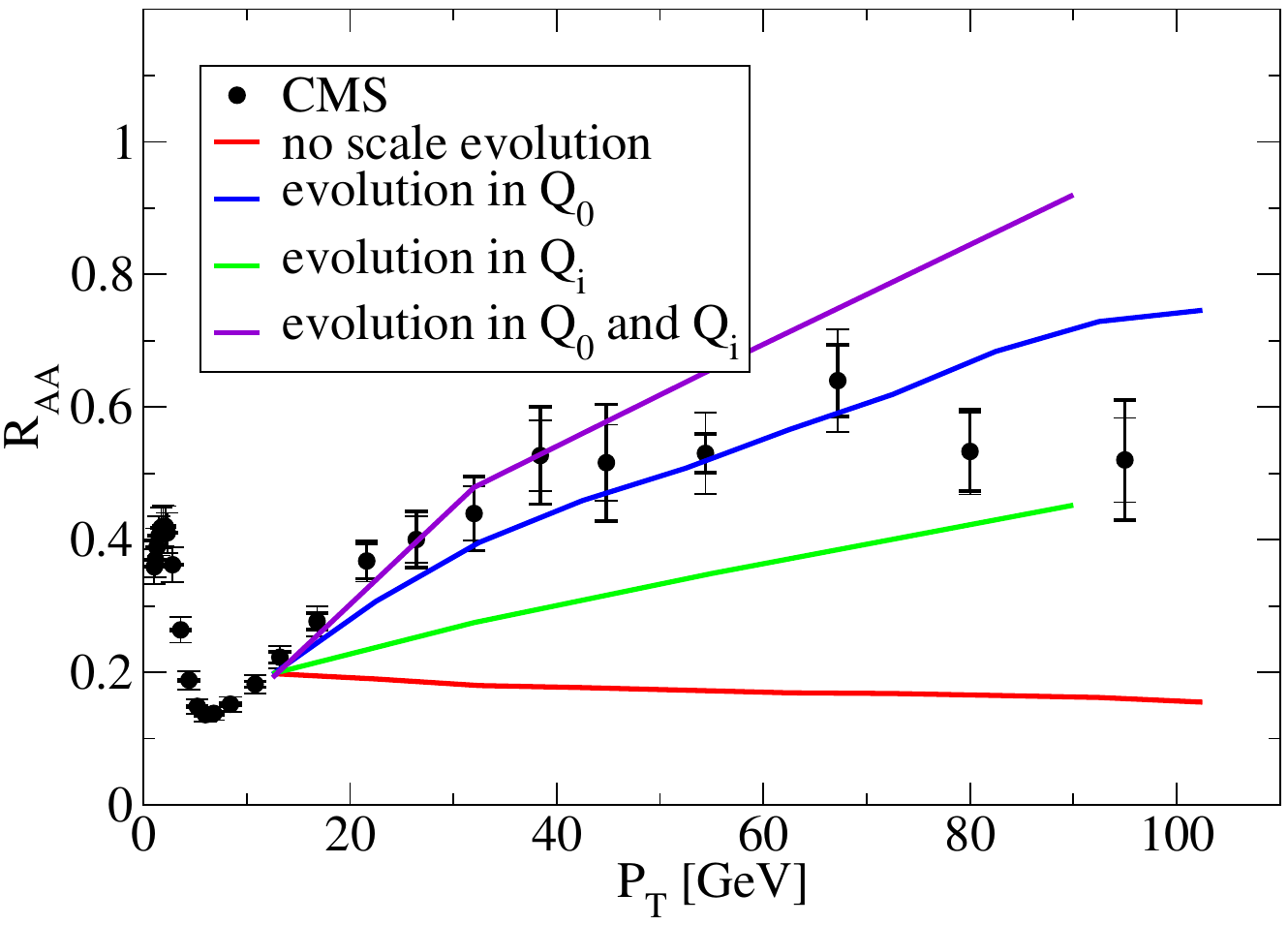, width=7cm} \epsfig{file=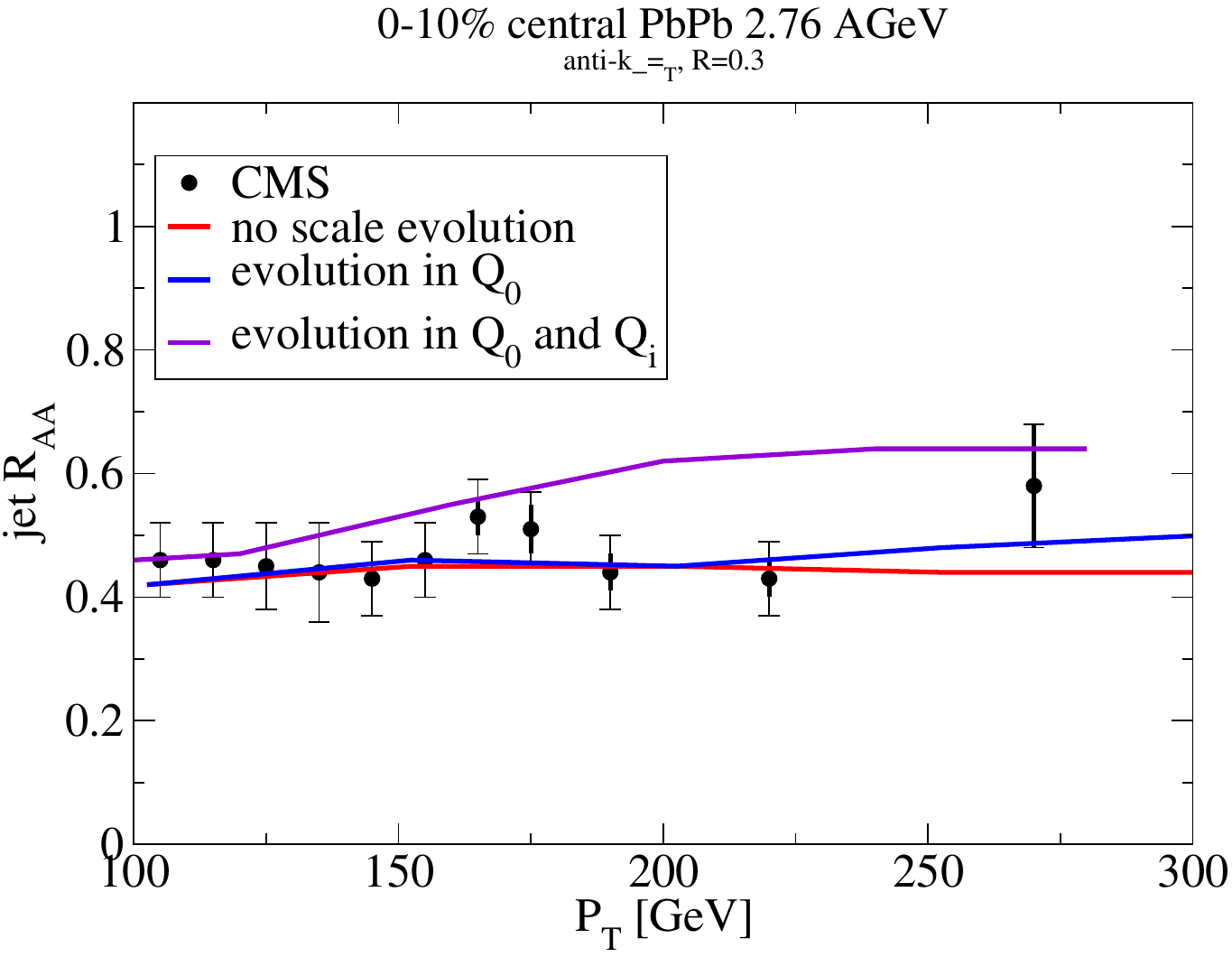, width=7cm}
\caption{\label{F-5}Nuclear suppression factor $R_{AA}$ of single hadrons (left) and jets (right) compared with various scenarios treating the QCD evolution in and out of the medium \cite{RAA-LHC} compared with CMS data \cite{CMS-RAA}.}
\end{center}
\end{figure}

This point is illustrated in Fig.~\ref{F-5} where the response of hadronic and jet $R_{AA}$ to different ways to treat the QCD scale evolution in- and out of the medium is shown \cite{RAA-LHC}. It can clearly be seen how clustering the observable decreases the ability of the observable to discriminate between different model scenarios.

\section{Current phenomenology}

The discussion of the preceding sections argues that maximal sensitivity to medium modifications can be obtained if an away side correlation is studied rather than a near side (to avoid the shower bias) and if the observable is not clustered. Indeed, h-h \cite{STAR-DzT} and jet-h \cite{STAR-jet-h} correlations paint a very accurate picture of medium-modified jets (seen under a particular trigger bias of course). Especially jet-h correlations, due to the high achievable statistics, image medium-modifications double-differentially longitudinal and transverse to the jet axis, and very few models \cite{jet-h} have reproduced this data set so far.

The picture of medium modifications seen in these observables is that above $\sim 3$ GeV, the longitudinal and transverse distribution of hadrons in the jet resemble vacuum results but are depleted. Below $\sim 3$ GeV, a broad and soft pedestal of hadrons formed from the medium-induced radiation is observed. The strength of this pedestal allows to constrain the fraction of direct elastic energy loss to medium degrees of freedom to about 10\% \cite{Elastic}. 

The same structure, filtered through a shower bias (cf. Fig.~\ref{F-2}) is observed e.g. by CMS in the jet fragmentation function analysis \cite{CMS-FF} at 100 GeV and above for LHC kinematics, and remarkably the scale of $\sim 3$ GeV at which the pedestal is observed is hardly changed, thus effectively ruling out any fractional energy loss mechanism (technically, the medium modification cannot be cast into the form of a modified splitting kernel $P(z) \rightarrow P'(z)$). The peculiar pattern of enhancement, depletion and return to unity in the fragmentation function ratio seen in the data is understood to result from the bias structure \cite{Bayesian,Abhijit}.

Through yet a different bias, the same structure is seen in $\gamma$-h correlations \cite{PHENIX-gamma-h}. Here, the fact that the channel $qg \rightarrow q\gamma$ is numerically much more important than $q \overline{q} \rightarrow g\gamma$ implies that the sample of jets recoiling from the photon is biased to be an almost pure sample of quark jets. This in principle allows to access the differences between quark and gluon jets. 

Jet correlations in which the observable is clustered such as the dijet imbalance \cite{ATLAS-AJ,CMS-AJ} do no longer show this structure explicitly but probe it implicitly. Many different theoretical models manage to reproduce the observable (see e.g. \cite{Muller,He,Young,YaJEM-AJ}), but as argued above and explicitly shown in \cite{YaJEM-AJ1}, a clustered observable is not strongly discriminating between different model assumptions.

Correlations of a photon with a jet ($\gamma$-jet, see e.g. \cite{CMS-gamma-jet}) again allow to access an almost pure quark jet sample, and are, to the currently limited statistics, currently also well described by models \cite{XN-gamma-jet,Dai}.

\section{Conclusions}

Through various forms of triggered correlation observables, we have achieved a differential understanding of how medium-modified jets look like at various energies, and there is a growing theoretical understanding how the filter of the particular bias of an observable changes how medium-modified jets appear in an analysis. In principle, these biases can be engineered to target selective physics questions, i.e. to study quark vs. gluon jets or to specifically probe certain regions of the medium. It would be fortunate if more efforts would be directed towards making this underlying picture more transparent --- on the experimental side by a consensus on what quantities and variables to plot to make results better comparable between experiments, and on the theory side by increased state of the art model comparisons with the most constraining rather than the least constraining measurements.

So far, all results of high $P_T$ physics are consistent with jet quenching being largely a kinematical broadening of the accessible radiation phase space, with some evidence for elastic energy transfer into the medium. This is of course in itself a non-trivial mechanism, as medium-induced gluon radiation is itself subject to further interaction with the medium, so the process has to be understood as a cascade transporting energy successively to lower energies and larger angles --- in other words, the jet in medium tends to thermalize. Currently there is no evidence that any observable requires a more interesting mechanism like breakdown of color coherence \cite{Coherence} or modifications to the color flow \cite{ColorFlow}, and possible dedicated observables probing specifically for these effects have to be engineered.

The converging understanding of the nature of the medium-modification implies that jet correlations might soon be utilized for the purpose they were originally envisioned, i.e. as well-calibrated probes to image the medium density evolution. Embedding medium-modifications to jets into state of the art event by event fluctuating hydrodynamics is feasible (see e.g. \cite{EbyE}), however has not resulted in a clear message so far --- there is a trend for such studies to prefer a higher initial eccentricity $\epsilon_n$ than hydrodynamics codes typically provide. Given that the pathlength dependence of the medium-modification is well tested in many correlation observables, this may perhaps point to an interesting gap in our understanding of the way the hard physics decouples from the fluid at low temperatures.

The new generation of tomographic observables that becomes feasible through a detailed understanding of the bias will go a long way in settling this and other interesting questions and finally establish hard probes as a quantitative tool to measure properties of the quark-gluon plasma.








\end{document}